\documentclass[a4paper]{amsart}

\usepackage{amsmath}
\usepackage{amsfonts}
\usepackage{amssymb}
\usepackage{graphicx}
\usepackage{color}

\newcommand{\bicross}{{\blacktriangleright\!\!\!\triangleleft}}

\newcommand{\lbiprod}{{>\!\!\!\triangleleft\kern-.33em\cdot}}
\newcommand{\rbiprod}{{\cdot\kern-.33em\triangleright\!\!\!<}}

\newcommand{\C}{{\Bbb C}}
\newcommand{\R}{{\Bbb R}}
\newcommand{\Z}{{\Bbb Z}}

\newcommand{\cg}{\mathfrak{g}}
\newcommand{\ch}{\mathfrak{h}}

\newcommand{\eps}{\epsilon}

\newcommand{\CM}{{\mathcal M}}

\newcommand{\tens}{\otimes}
\newcommand{\id}{{\rm id}}
\newcommand{\extd}{{\rm d}}

\newcommand{\<}{{\langle}}

\renewcommand{\>}{{\rangle}}

\begin{document}


\author{Shahn Majid}
\address{School of Mathematical Sciences\\ Queen Mary University of London \\ Mile End Rd, London E1 4NS }



 \email{s.majid@qmul.ac.uk}

\title{Quantum Gravity: are we there yet?}

\begin{abstract} The turn of the millennium was a time of optimism about an approach to noncommutative geometry inspired by rich mathematical objects called `quantum groups' and its applications to quantum spacetime. This would model  quantum gravity effects  as noncommutativity of spacetime coordinates and was arguably going to solve quantum gravity itself. It took a further 20 years from that point to develop a particularly suitable formalism of `quantum Riemannian geometry', but this was largely done and has begun to be used to construct baby quantum gravity models. In this article, we obtain new results for state of the art fuzzy sphere and $n$-gon models in this approach. We also review what are some elements of quantum gravity that we can already see and what are the critical conceptual and mathematical elements that are still missing to more fully achieve this goal. 
\end{abstract}

\maketitle

\section{Introduction} 

While the start of the 20th century told us that position and momentum coordinates of mechanical systems are better modelled by quantum theory as operators that do not commute (the famous Heisenberg uncertainty relations), the tail end of the 20th century saw  emergence of the view that spacetime coordinates themselves might need to not commute, this time due to quantum gravity corrections to geometry itself. This is the {\em Quantum Spacetime Hypothesis} and one of the first (and still the most widely studied) models of it was the bicrossproduct spacetime\cite{MaRue}
\begin{equation}\label{bicmodel} [x^i,t]=\imath\lambda_P x^i,\quad [x^i,x^j]=0\end{equation}
where $x^i=x,y,z$ are spatial coordinates, $t$ is time and $\lambda_P\sim 10^{-44}$ seconds is the Planck time, an extremely small parameter. The idea is that there is lots of evidence that spacetime at this sort of timescale or, equivalently, the
Planck length scale $10^{-35}$m, is not a continuum but something else, indeed a deep mystery in the absence of a theory of quantum gravity.  The quantum spacetime hypothesis is a concrete proposal for how that `something else' could be better modelled, even without knowing quantum gravity, and hence a better foundation on which to build it.

You might complain at this point that surely a model like (\ref{bicmodel})  depends on the reference frame, i.e. breaks Einstein's principle of Special Relativity? Actually what happens is that the entire Poincar\'e group of translations and Lorentz transformations is also modified, using the notion of a {\em quantum group}. This was a generalisation of group theory that had emerged in the 1980s (although the axioms, but without convincing examples, were proposed by the mathematician H. Hopf in the 1940s). One class of these, the q-deformations\cite{Dri} came out of integrable systems and another class, the bicrossproducts\cite{Ma:pla} came out of the search for models of quantum gravity. The relevant quantum group $\C[H^{1,3}]\bicross U(so_{1,3})$ for the above model is one of the latter group, albeit a quantum group isomorphic to it (without an action on a quantum spacetime) was first proposed in \cite{Luk} as a contraction limit of $U_q(so_{2,3})$, i.e. the two are not unconnected. You also might complain that as $\lambda_P$ is so small, isn't this model totally untestable? However, there are situations where the quantum-gravity effect can be magnified and (in principle) measurable. For example, a predicted small energy-dependence on the speed of light in the above model would result in different arrival times of a few milliseconds for $\gamma$-ray bursts of cosmological origin\cite{AmeMa}. Other effects could be a phase transition such as `Bose-Einstein condensation' (albeit, so far modelled in mathematically analogous situations rather than quantum gravity itself). There were also models proposed on other theoretical grounds, such as in string theory as an effective model of the motion of the ends of open strings landing on a d-brane\cite{SeiWit}, or a model in \cite{DFR} keeping Lorentz symmetry. It was also possible to see concrete models emerging in Euclideanised 2+1 quantum gravity, which is exactly solvable but topological (i.e. without dynamics of the graviton), namely the `fuzzy $\R^3$' model,
\begin{equation}\label{spinmodel}  [x^i,x^j]=2\imath\lambda_P \eps_{ijk}x^k\end{equation}
where $\lambda_P$ is now of order the Planck length and $\eps_{123}=1$ is a totally antisymmetric tensor. This was proposed in \cite{Hoo}  but the reader may be familiar with it in another context as the algebra of angular momentum, i.e. the enveloping algebra $U(su_2)$. This has a quantum Poincar\'e group $\C(SU_2)\rtimes U(su_2)$ as a special case (a `quantum double') of a bicrossproduct\cite{MaSch,FreMa}. 

If such models of flat but quantum spacetime were the state of the art around the turn of the millennium, where did they need to go next, and where did they go next? The bottom line is that while they motivated speculative ideas for  `quantum gravity phenomenology', they still needed two fundamental issues to be addressed:

\begin{enumerate}
\item How to extend such models to quantum but curved spacetimes such as around a black hole or in cosmological models?
\item How to extract physical predictions from the mathematics in a coordinate-free way?
\end{enumerate}

Neither is an easy task and it has taken the next 20 years, i.e. up until now, to address them in a practical manner. On the first problem, powerful approaches to noncommutative geometry already existed, notably that of Alain Connes\cite{Con,ChaCon} and,  among other things, encoded geometric information in spectral data (notably via an abstract `Dirac operator'). But this did not give direct access to analogues of familiar geometric objects such as the metric tensor, and was also not typically well adapted to examples coming from quantum groups.  Instead, what emerged was a `layer by layer' approach dubbed {\em Quantum Riemannian Geometry} (QRG), in which we start with a potentially noncommutative coordinate algebra $A$, chose its differential structure in the sense of a bimodule of 1-forms $\Omega^1$, chose a metric $\cg\in\Omega^1\tens_A\Omega^1$ and solve for a {\em quantum Levi-Civita connection} (QLC) $\nabla:\Omega^1\to \Omega^1\tens_A\Omega^1$, with Riemann curvature $R_\nabla$. This is now covered in my text with Beggs\cite{BegMa} building on key works such as \cite{DVM,BegMa:gra, BegMa:rie}. The state of the art here is that while $R_\nabla$ is canonical, there is only a `working definition' of the Ricci curvature. But this is enough to build first `baby quantum gravity' models\cite{Ma:squ,Ma:haw, LirMa1,ArgMa1} and get a first look at quantum gravity. 

The second problem is equally fundamental. In General Relativity,   calculations can be done in any coordinate system, but it is then very hard to know what aspects of what one computes are due to the choice of coordinates and what is actually physical, independent of the coordinates. Key here is the notion of a geodesic as a way to map out the geometry of a continuum spacetime in a coordinate-invariant way by looking at how particles move in it. The same issues arise for quantum spacetime not only on the geometric side but even in flat spacetime, i.e. working with specific generators and relations, how can we extract the actual physics? This was recently addressed in the notion of `quantum geodesics' \cite{Beg:geo, BegMa:geo, BegMa:cur, LiuMa1}.

Before we get started, here is a current wish-list of things that we might hope that quantum gravity could shed light on.
\begin{enumerate}
\item[{\rm Q1}] What happens at the centre of a black hole or at the initial `Big Bang' where, classically, the curvature diverges? 
\item[{\rm Q2}] Can we explain the problem of the cosmological constant? To match observed cosmology to Einsteins gravity, one need to add an extra `cosmological constant' term or, equivalently, an unexplained `dark energy' to the stress tensor of a vacuum. The energy density needed is of order $10^{-29}$g/cm${}^3$ and the problem is why, if it is a quantum gravity effect, is it so small compared to the Planck density and yet nonzero?
\item[{\rm Q3}] Can we more deeply understand an apparent link between curvature and entropy as suggested by Bekenstein-Hawking radiation near a black-hole? 
\item[{\rm Q4}]  Can we understand the Diosi-Penrose idea of `gravitational state reduction' where quantum systems are proposed to spontaneously measured due to interaction with gravity?
\item[{\rm Q5}] Can we understand the structure of the Standard Model, i.e. the particular elementary particles observed in accelerators, and their particular masses, as emerging from quantum gravity?
\end{enumerate}

I won't have space here to discuss these in any depth and nor to review or connect with other approaches to quantum gravity, instead focussing on the QRG approach. We first provide a brief outline of the QRG formalism, which readers who do not like algebra should just skip over. Section~\ref{secbaby} then contains the new results of the paper, namely an in-depth further study of two baby quantum gravity models and not published elsewhere. Section~\ref{secKK} and Section~\ref{secgeo} overview some other key topics and we conclude in Section~\ref{secrem} with how far along we are on the wish-list and what are the main obstructions we need next to overcome.

\section{QRG formalism}\label{secqrg} A lightning introduction to the formalism is as follows. 

(i) We pick a $*$-algebra $A$ in the role of complexified coordinates on a continuum manifold, i.e. equipped with a complex linear map $*:A\to A$ reversing products and squaring to the identity. In the commutative case one could look at the real subalgebra of self-adjoint elements (which in the classical case would be the usual real coordinate algebra) but in the noncommutative case one just remembers $*$ as specifying the `real form'. In quantum mechanics $A$ could be a matrix algebra then $*$ is just hermitian conjugation. 

(ii) We pick a graded algebra $\Omega=\oplus_i\Omega^i$ in the role of differential forms of degree $i$. This replaces `differential calculus' in classical geometry and as such is equipped with $\extd: \Omega^i\to \Omega^{i+1}$ obeying a graded-Leibniz rule. We assume $\Omega$ is generated by $\Omega^0=A$ and $\Omega^1=A\extd A$ and that $*$ extends to $\Omega$ as a graded-involution commuting with $\extd$. The product of differential forms is denoted $\wedge$. 

(iii) We pick a quantum metric $\cg\in \Omega^1\tens_A\Omega^1$. Although geometers do not think algebraically, this is actually what they are doing when they write a metric as $\cg=g_{\mu\nu}\extd x^\mu \extd x^\nu$ in local ordinates on a manifold $M$, where $\{\extd x^\mu\}$ are a basis of 1-forms and there is a hidden $\tens_{C^\infty(M)}$ in the middle of $\extd x^\mu \extd x^\nu$ to be understood. We need $\cg$ to be invertible in the sense of a map $(\ ,\ ):\Omega^1\tens_A\Omega^1
\to A$ which classically would be given by the inverse metric tensor $g^{\mu\nu}=(\extd x^\mu,\extd x^\nu)$. (In mathematical terms $\Omega^1$ is an $A$-bimodule and a quantum metric makes it isomorphic to its dual in the monoidal category of $A$-bimodules.) We usually impose some form of `quantum symmetry' in addition, such as $\wedge(\cg)=0$. We also require `reality' in the form 
\[ \dagger(\cg)=\cg,\quad \dagger={\rm flip}(*\tens *),\]
which in local coordinates in the classical case would ensure that the coefficient in a self-adjoint basis are real. 

(iv) A Riemannian connection is formulated as a map $\nabla:\Omega^1\to \Omega^1\tens_A\Omega^1$ with certain properties. Note that a `right vector field' is a map $X:\Omega^1\to A$ which respects right-multiplication by $A$ and in this case $\nabla_X=(X\tens
\id)\nabla$ is the associated `covariant derivative' familiar in the continuum case. The connection map is required to obey
\[ \nabla(a\omega)=\extd a\tens\omega+a\nabla\omega,\quad \nabla(\omega a)=\sigma(\omega\tens \extd a)+ (\nabla\omega)a\]
for some bimodule map $\sigma:\Omega^1\tens_A\Omega^1\to \Omega^1\tens_A\Omega^1$, which we assume invertible. The latter in classical geometry would be an invisible `flip' of indices needed for a vector field $X$ to couple correctly on the left-most copy of $\Omega^1$. It is not additional data but, if it exists, determined by $\nabla$ via the above. The category of such `bimodule connections' is closed under tensor product, so $\nabla$ extends to a connection on $\cg\in \Omega^1\tens_A\Omega^1$ and we say it is {\em metric compatible} if $\nabla(\cg)=0$. There is also a torsion tensor of any connection 
\[ T_\nabla=\wedge\nabla-\extd: \Omega^1\to 
\Omega^2\]
comparing the exterior derivative $\extd:\Omega^1\to \Omega^2$ with $\nabla$ followed by $\wedge$. A {\em quantum Levi-Civita connection} (QLC) is defined as a $\nabla$ which is metric compatible and has $T_\nabla=0$. Unlike in classical geometry, it need not exist and even if it does, it need not be uniquely determined by $\cg$. It means that not every classical metric is quantisable to a QRG as we currently formulate it, which we see as a good thing as it may single out particular classical metrics as arising from QRG (see Section~\ref{secKK}). We also require `reality' or {\em $*$-preserving} in the sense
\[ \sigma^{-1}\circ \nabla\circ *= \dagger\nabla.\]
In the classical case, this ensures that the Christoffel symbols or connection coefficients with respect to a self-adjoint basis are real. 

(v) The Riemann curvature of a connection is a map 
\[ R_\nabla=(\extd\tens\id-\id\wedge\nabla)\nabla:\Omega^1\to \Omega^2\tens_A\Omega^1.\]
Both this and $T_\nabla$ are well-defined for any left connection, i.e. not using the Leibniz rule involving $\sigma$. This `algebraic' formulation of geometry will be very alien to readers more familiar with General Relativity, where formulae are generally given in local coordinates and tensor calculus, but in many ways is cleaner and more conceptual. 

The Ricci tensor is less clear and needs some kind of interior product $\Omega^1\tens_A\Omega^2\to \Omega^1$ in order to be able to trace $R_\nabla$. The simplest additional data here is to suppose a bimodule lifting map $i:\Omega^2\to \Omega^1\tens_A \Omega^1$, which in classical geometry simply writes a 2-form as an antisymmetric combination of 1-forms (i.e. with two, antisymmetric, indices). In QRG we specify it so that $\wedge \circ i=\id$ and ideally $i\circ *=-*\circ i$, as this would be true in classical geometry. Then
\[ {\rm Ricci}=((\ ,\ )\tens \id)(\id\tens i\tens \id)(\id\tens R_\nabla)\cg \in \Omega^1\tens_A\Omega^1,\]
after which we define the {\em Ricci scalar curvature}
\[ R=(\ ,\ ){\rm Ricci}.\]
This is needed for the action for quantum gravity. But note that the natural QRG Ricci {\em reduces in the classical case to $-{1\over 2}$ of the usual value}. There also a natural QRG Laplacian $\Delta=(\ ,\ )\nabla\extd$ on $A$ which provides the action for a scalar field on the QRG.

\section{Baby quantum gravity models}\label{secbaby}

To write down quantum gravity on a QRG, we need two more things. There is usually a natural choice for each of these but no fully general theory, i.e. they have to be chosen for each model. One is a notion of integration $\int: A\to \C$ which sends non-zero positive  elements to $\R_{>0}$. The second is, even if we can solve for the moduli $\CM_{QRG}$ of QRG's for a given differential algebra $(A,\Omega^1,\Omega^2,\extd)$, we need to choose a measure $\mu_{\cg}$ for integration over this classical moduli space.   Then we can write the quantum gravity partition function
\begin{equation}\label{QG} Z=\int_{\CM_{QRG}} \extd \cg\  \mu_{\cg}\,  e^{{\imath \over G}\int R},\end{equation}
where $G$ is a real coupling (say, positive) constant. We are suppressing here that there may also be parameters in the QLC (it need not be unique in QRG) and these should also be either fixed as background data or integrated over. For Euclideanised quantum gravity,  we replace the imaginary unit $\imath$ in the exponent by $\pm 1$ according to which way $\int R$ is bounded.

\subsection{Quantum gravity on the fuzzy sphere revisited} 

The unit fuzzy sphere has self-adjoint generators $x^i$ with relations
\[ [x^i,x^i]=2\imath\lambda_P \eps_{ijk}x^k,\quad \sum x^i{}^2=1-\lambda_P^2\]
where $\lambda_P$ is a dimensionless parameter. This is equivalent to a sphere in (\ref{spinmodel}) except that we divided through by the size of the sphere (one could scale the $x^i$ and $\lambda_P$ by this to work with physical dimensions). There is a natural differential calculus\cite{BegMa}, cf. \cite{Mad}, with 3 self-adjoint central basis 1-forms $s^i$. These correspond classically to the Killing forms for the action of $SU_2$, under which the fuzzy sphere and calculus are covariant (classically, these do not form a basis but this is not the case for $\lambda_P\ne 0$). A metric is therefore of the form
\[ \cg=g_{ij} s^i\tens s^j\]
for a positive real symmetric matrix $\{g_{ij}\}$. The Ricci scalar works out \cite{LirMa1} as
\[ R={1\over 2 \det(g)}\left({\rm Tr}(g^2)-{1\over 2}{\rm Tr}(g)^2\right)\]
We proceed as in \cite{LirMa1} and take action with measure $\int 1= \det(g)$ and make a spectral decomposition to eigenvalues $\lambda_1,\lambda_2,\lambda_3>0$ and Euler angles. In these terms, the Ricci scalar is
\[ R={1\over 4\lambda_1\lambda_2\lambda_3}\left(\lambda_1^2+\lambda_2^2+\lambda_3^2-2(\lambda_1\lambda_2+\lambda_2\lambda_3+\lambda_3\lambda_1)   \right).\]
If we are only interested in observables which are functions of the eigenvalues then the effective partition function for Euclideanised quantum gravity is
\begin{equation}\label{Zfuz}Z=\int_\eps^L \prod_i \extd\lambda_i\, {|(\lambda_1-\lambda_2)(\lambda_2-\lambda_3)(\lambda_3-\lambda_1)|\over \lambda_1^2\lambda_2^2\lambda_3^2}\, e^{-{1\over 2G^2}(\lambda_1^2+\lambda_2^2+\lambda_3^2-2(\lambda_1\lambda_2+\lambda_2\lambda_3+\lambda_3\lambda_1))},\end{equation}
where we cut-off at both ends to $\eps<\lambda_i<L$. The lower end, while divergent,  still has a negligible contribution compared to that from the other end at large $L$, for machine-readable $\eps>0$. Unless stated otherwise, we use $\eps=10^{-20}$ as default, but our graphs, for example, do not look significantly different all the way up to $10^{-1}$.  The physical values of $L,\eps,G$ have dimensions length${}^2$ and it is for this reason that we squared $G$ compared to \cite{LirMa1} and the general scheme. 

\subsection{Higher order corrections for large $L$}

Henceforth, we in any case set $G=1$, either because we are working in Planckian units or because we can just use rescaled variables and parameters
\[ \lambda_i={\lambda_i^{phys}\over G},\quad L={L^{phys}\over G},\quad \eps={\eps^{phys}\over G}\]
where $\lambda_i^{phys},L^{phys},\eps^{phys}$ are what we previously called $\lambda_i,L,\eps$. In terms of the new dimensionless variables, $Z$ looks identical, but with $G=1$.  Our first new result is that
 \[ Z\sim {1 \over  4 L^{12}} e^{3 L^2\over 2} \]
for $L>>6$, with increasing accuracy as $L\to\infty$ (and any sufficiently small $\eps$). This is already a very good approximation for $L>30$. 

\begin{figure}
\[ \includegraphics[scale=0.7]{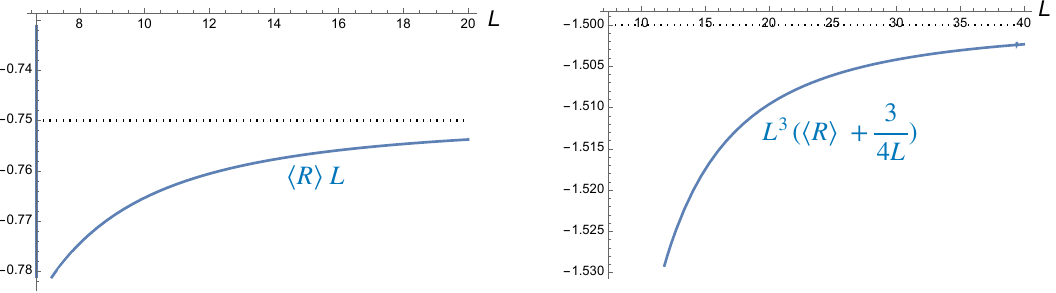}\]
\caption{Expectation value of Ricci scalar scalar showing rapid convergence of $\<R\>L$ to $-3/4$ and first quantum gravity correction. \label{figfuz}}
\end{figure}

Next, for expectation values, we insert an operator into the integral for $Z$ and divide the result by $Z$. Then for large $L$, one has
\[ \<\lambda_{i_1}\cdots \lambda_{i_n}\>\sim  L^n\]
as $L\to\infty$, at least for positive powers. This is a correction to \cite{LirMa1}, where there was a constant coefficient $3\over 16$ due to a coding error in the calculation there. Our new result here is that to leading order 
\begin{equation}\label{Deltafuz} \Delta \lambda_i := \sqrt{\<\lambda_i^2\>-\<\lambda_i\>^2}\sim \sqrt{41\over 12}{1\over L}\end{equation}
as $L\to \infty$, as can be deduced from the more detailed expansion
\[\< \lambda_i\>= L - {25\over 6 L}- {1\over 18 L^3}+ O({1\over L^4}),\quad \<\lambda_i\lambda_j\>=\begin{cases}L^2- {17\over 3}+ {7\over 3 L^2}+O({1\over L^3}) & i\ne j\\ 
L^2- {25\over 3}+{62\over 3 L^2}+O({1\over L^3}) & i=j\end{cases}.\]
Here, (\ref{Deltafuz})  says that the  theory does exhibit quantum gravity (there is quantum uncertainty in the metric values) but less so at large $L$, which makes sense as this would be more dominated by macroscopic behaviour. Similarly, one has 
\[ \< R\>=- {3\over 4 L}-{3\over 2 L^3}+ O({1\over L^4})\]
as shown in Figure~\ref{figfuz}. The leading term here is (in QRG conventions) the Ricci curvature of a unit fuzzy sphere with metric $g_{ij}=L\delta_{ij}$, in other words the classical value. The next term is then the leading quantum gravity correction. 

\subsection{Renormalisation of the gravitational coupling constant $G$} We can translate these results back to the physical values defined via $\lambda^{phys}_i$. The Ricci curvature computed from the latter is 
\[ R^{phys}={R\over G},\quad \<R^{phys}\>={\<R\>\over G}\sim -{3\over 4 \<\lambda_i\> G}=-{3\over 4 \<\lambda_i^{phys}\>}\]
for $L\to\infty$. We do not need it if we only look at ratios of physical quantities, but we can also 
consider $G=G(L)$ and choose this so that 
\[ \<\lambda^{phys}_i\>:=G(L_0)\<\lambda_i\>_{L_0}\]
is fixed as some arbitrary value at some value $L_0$ of the regularization parameter $L$. The subscript is used to indicate our above computations done at $L_0$. Then one can in principle solve for $G(L)$ in such a way 
\[ \<\lambda^{phys}_i\>=G(L)\<\lambda_i\>_{L},\]
i.e. the right hand side is constant for all $L$. This gives
\[ G(L)= {\<\lambda^{phys}_i\>\over \<\lambda_i\>_L}=G(L_0){ \<\lambda_i\>_{L_0}\over  \<\lambda_i\>_L}\sim {G(L_0)L_0\over L}(1-{25\over 6 L_0^2})(1+ {25\over 6 L^2})\]
for large $L_0,L$  by our above. Using more precise information about $\<\lambda_i\>_L$, one can get more precise  information on  $G(L)$.  We can then then compute 
\[ \<R^{phys}\>_L={ \<R\>_L\over G(L)}=\<R\>_L {\<\lambda_i\>_L\over \<\lambda^{phys}_i\>}\sim \left(-{3\over 4}+{13\over 8 L^2}+ O({1\over L^3}) \right){1\over \<\lambda^{phys}_i\>}\]
for large $L$. The value here depends on the regulator $L$ but to lowest order in large $L$, we can replace this by  $\<\lambda^{phys}_i\>/G$ so that in physical terms
\[ \<R^{phys}\>\sim \left(-{3\over 4}+{13 G^2\over 8 \<\lambda^{phys}_i\>^2 }\right){1\over \<\lambda^{phys}_i\>}\]
plus higher order corrections. Here the LHS and $G$ are as measured using cut-off $L$. As $L\to\infty$, we get the previous macroscopic value but we also now see the first quantum corrections. 

\subsection{Phase transition at $L\approx 6$ to the deep quantum gravity limit}

\begin{figure}
\[ \includegraphics[scale=.75]{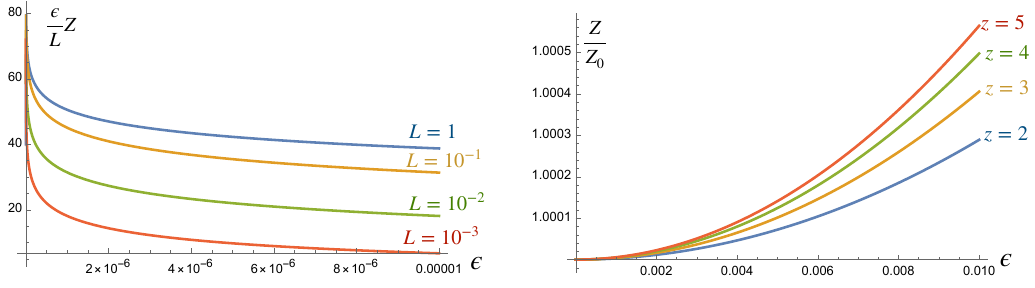}\]
\caption{Numerical behaviour of $Z$ for $L<<6$ and  $Z/Z_0$ for fixed $z={L\over\eps}>1$ showing convergence to 1 as $\eps\to 0$. \label{figZeps}}
\end{figure}

On the other hand, as we decrease $L$, the value of $Z$ suddenly jumps to a very high value of order $1/\eps$, at around $L=6.07$ for small enough $\eps$  (otherwise there is a slightly earlier transition point, for example at around $L=5.5$ for $\eps=10^{-5}$). Since Mathematica in this regime  shows significant non-convergence warnings in the numerical integration,  the precise values here and in Figure~\ref{figZeps} should be regarded as indicative. Nevertheless, there is clearly a very different behaviour for $L$ below the critical value, better approximated by something like
\[ Z\sim c(\eps,L) {L\over\eps}\ln({L\over\eps})\]
as $\eps\to 0$, with some residual dependence in the coefficient $c$. This indicates a phase transition from a macroscopic $L$-dominated regime to a Planckian $\eps$-dominated regime. The results in the latter phase being dominated by small $\eps$, one can drop the exponential factor in the integrand, i.e.
\[ Z\approx Z_0:= \int_\eps^L \prod_i \extd\lambda_i\, {|(\lambda_1-\lambda_2)(\lambda_2-\lambda_3)(\lambda_3-\lambda_1)|\over \lambda_1^2\lambda_2^2\lambda_3^2}\]
for  $L<<1$. The ratio of the two as we vary $z:={L\over\eps}$ is also shown in the figure. This $Z_0$ is independent of scaling the $\lambda_i$, so we can also view it as the original partition function (\ref{Zfuz}) at $G=\infty$ before we rescaled fields and with our current limits $L,\eps$ regarded as the physical $L^{phys},\eps^{phys}$ in this interpretation. Hence this is the {\em deep quantum gravity} limit of the theory and is of interest in its own right. 

\begin{figure}
\[ \includegraphics[scale=.75]{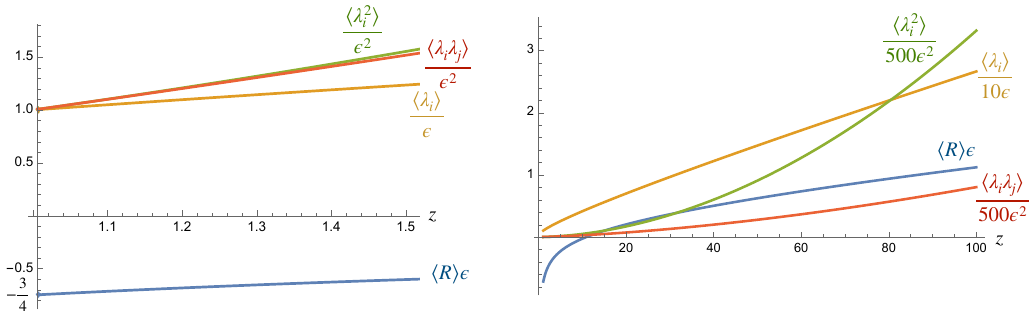}\]
\caption{Expectation values in the deep quantum gravity $Z_0$ phase of the fuzzy sphere as a function of small and large $z$ respectively. \label{figfuzeps}}
\end{figure}

Moreover, the $Z_0$ theory can be computed exactly as follows and necessarily depends only on the ratio $z:={L\over\eps}$.  If $\lambda_1>\lambda_2$ then there are three regions for $\int\extd\lambda_3$ in relation to these, each of which can be done analytically, which we then add together. If $\lambda_2>\lambda_1$ then we obtain the same as the other way around. This gives us
\[ Z_0=\int_\eps^L{\extd\lambda_1\extd\lambda_2 \over\lambda_1^2\lambda_2^2} f(\lambda_1,\lambda_2)\]
\[ f(\lambda_1,\lambda_2)= |\lambda_1-\lambda_2|\left(L-\eps- 4|\lambda_1-\lambda_2|+ ({1\over\eps}-{1\over L})\lambda_1\lambda_2+  (\lambda_1+\lambda_2)(2|\ln({\lambda_1\over\lambda_2})|-\ln({L\over \eps})\right).\]
We then have two regions for $\int\extd\lambda_2$ in relation to $\lambda_1$, each of which can be done analytically and the results added, to give a function of $\lambda_1$. This too can be integrated analytically, to give 
\[ Z_0={6\over  z} \left(\ln (z )(z^2-1)+2 z \ln( z   ) -4 (z-1)^2\right).\]
One similarly has exact expressions for expectation values in this phase, where we insert these observables into the integral for $Z_0$ and then divide by $Z_0$. These are respectively 
\begin{align*}  \<\lambda_i\>&= {\eps (z-1)\over Z_0}\left((1+4z+z^2)\ln (z ) -3 (z^2-1)\right),\\
\<\lambda_i^2\>&={\eps^2 (z-1)\over 9 Z_0}\left(6(1+z)(1+5z+z^2)\ln (z ) -(z-1)(19+46z+19z^2)\right)\\
\<\lambda_i\lambda_j\>&={\eps^2 (z-1)\over 3 Z_0}\left(6z(1+z)\ln (z ) -(z-1)(1+10z+z^2)\right)\\
\<R\>&={(z-1)\over 8 \eps z^2 Z_0}\left(6z(3+10z+3z^2)\ln (z ) -(z^2-1)(1+46z+z^2)\right)\end{align*}
for $i\ne j$ and are plotted in Figure~\ref{figfuzeps} as functions of $z$. We see a limiting value
\[ \<\lambda_i\>=\eps,\quad \<\lambda_i^2\>=\<\lambda_i\lambda_j\>=\eps^2,\quad \<R\>=-{3\over 4\eps}\]
as $z\to 1$, which has classical behaviour with zero variance of the $\lambda_i$ and Ricci curvature that of a fuzzy sphere with metric $g_{ij}=\eps\delta_{ij}$. As $z$ increases all, the expectation values increase, with the curvature crossing zero at $z\approx 10.88115$. The asymptotic form for large $z\to\infty$ is by contrast
\[   \<\lambda_i\>\sim {\eps z\over 6},\quad \<\lambda_i^2\>\sim {\eps^2 z^2 \over 9},\quad \<\lambda_i\lambda_j\>\sim {\eps^2 z^2\over 18 (\ln(z)-4)},\quad \< R\>\sim {z\over 48\eps( \ln(z)-4)}\]  
where we see highly non-classical behaviour. This limit of large $z$ form implies for the deep quantum gravity theory at $G=\infty$ defined by $Z_0$ and fixed $L$ that 
\[ \<\lambda_i\>\to {L\over 6},\quad \<\lambda_i^2\>\to {L^2\over 9},\quad  \<\lambda_i\lambda_j\>\sim {L^2\over 18 \ln({1\over\eps})},\quad \< R\>\sim {L\over 48\eps^2 \ln({1\over\eps})} \]
as $\eps\to 0$. From the two points of view, we also have
\[ {\Delta\lambda_i\over\<\lambda_i\>}\sim  \sqrt{3},\quad {\<\lambda_i\lambda_i\>\over\<\lambda_i\>\<\lambda_j\>} \sim {2\over \ln(z)-4}\sim {2\over\ln({1\over\eps})},\quad\<R\>\<\lambda_i\>\sim {1\over 8 \eps^2 (\ln(z)-4)}\sim {1\over 8 \eps^2 \ln({1\over\eps})}\]
which shows a fixed relative uncertainty in the quantum metric fields $\lambda_i$ similarly to the deep quantum gravity limits of other known models. However, we also see $\<\lambda_1\lambda_2\>$ going to zero and the Ricci curvature diverging as the regulator $\eps\to 0$. This gives a flavour of this phase of the quantum gravity model as fundamentally different from the large $L$ phase studied before. In this phase, we would renormalise the original $Z$ theory with $G$ a function of $\eps$ or $z$ rather than $L$. 

\subsection{Quantum gravity models on graphs}

\begin{figure}
\[ \includegraphics[scale=.8]{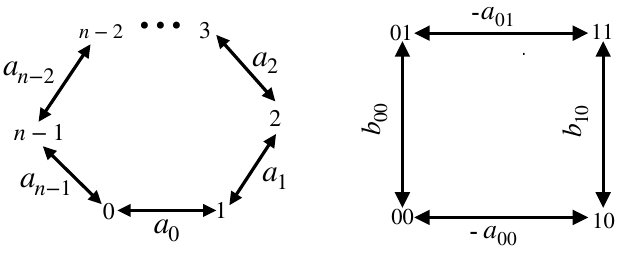}\]
\caption{For the QRG of a graph, arrows are differential forms and a quantum metric is an assignment of a non-zero real `square length' to each edge. The quantum metric for the  $n$-gon defines a function $a$ which at vertex $i$ is $a_i$, and similarly for the square, two functions $a,b$  with $a_{10}=a_{00}, a_{11}=a_{01}$ and $b_{01}=b_{00}, b_{11}=b_{10}$. \label{figsquare}}
\end{figure}

QRG can be applied to any algebra but it includes as a special case the algebra $A$ of functions in a discrete set $X$. Here, possible $\Omega^1$ are in 1-1 correspondence with directed graphs with vertex set $X$. A vector space basis of 1-forms is $\{\omega_{x\to y}\}$ labelled by the arrows and $\extd f=\sum_{x\to y} (f(y)-f(x))\omega_{x\to y}$ for $f\in A$ encodes finite differences across each arrow. Multiplication of 1-forms by functions is $f\omega_{x\to y}=f(x)\omega_{x\to y}$ and $\omega_{x\to y}f=f(y)\omega_{xto y}$, i.e. noncommutative even though the algebra $A$ itself is commutative. In this way, a directed graph is literally an example of quantum geometry. A metric here, in order to have a bimodule inverse, has the form
\[ \cg=\sum_{x
\to y}g_{x\to y}\omega_{x
to y}\tens\omega_{y\to x},\quad g_{x\to y}\in \R_{\ne 0}\]
and only exists if every arrow has a reverse arrow, which we assume. In principle the `metric square-length' of an arrow could depend on the direction, but a natural notion of symmetry here is to suppose that it does not. We proceed in this edge-symmetric case, albeit this is not the only case of interest. Then a quantum metric is nothing other than an assignment of a nonzero real number or `square-length' to every edge. Examples are shown in Figure~\ref{figsquare} for a Euclidean $n$-gon for $n\ge 3$ and a Lorentzian square, so called because the horizontal metric weights are assumed negative or `spacelike' (with $a_{00}, a_{01}>0$) while the vertical ones are `timelike' with $b_{00}, b_{10}>0$).   For the Euclidean case,  the usual convention is to take all the $a_i>0$.

\begin{figure}
\[\includegraphics[scale=0.75]{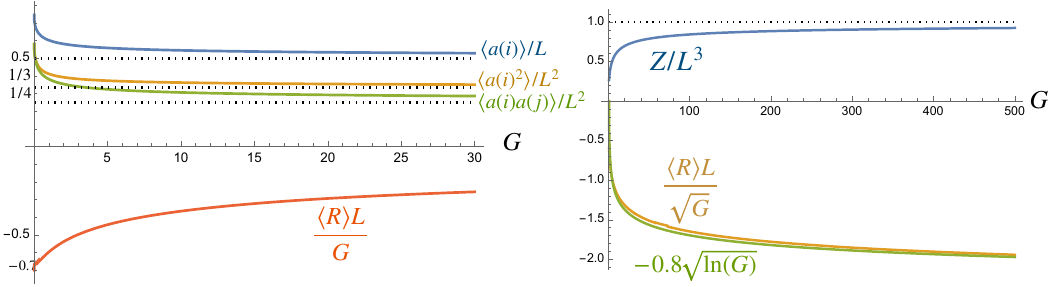}\]
\caption{\label{Z3figs} $G$-dependence of vacuum expectation values and partition function for quantum gravity on a triangle.}
\end{figure}

Proceeding in the $n$-gon case, there is a basis $e^\pm$ over the algebra of left-invariant 1-forms with respect to the group $\Z_n$,  which we take to anticommute for the wedge product. Then the metric values regarded as a function $a$ on the vertices provides the metric tensor in the form 
\[ \cg=a e^+\tens e^- + R_-(a) e^- \tens e^+\]
where $R_-(a)(i)=a(i-1)$ a shift mod $i$. There is a natural QLC which is unique for $n\ne 4$ (we stick to this one for all $n$)  and given in \cite{Ma:haw,ArgMa1}. We  omit the details and jump to the Ricci scalar
\[ R(i)=-{1\over 2}\left({ a(i-1)\over a(i)^2}+ {a(i)\over a(i-1)^2}-{1\over a(i+1)}-{1\over a(i-2)}\right)\]
defined with respect to an obvious antisymmetric lift map $i$. We let the measure $\mu$ for integration be $a$ itself as a natural choice.Then the quantum gravity partition function for $n=3$, say, is
\[ Z=\int_0^L\int_0^L\int_0^L\extd a(0)\extd a(1)\extd a(2)\, e^{{1\over 2G}\left( {a(2)\over a(0)}+ {a(0)\over a(1)}+{a(1)\over a(2)}- {a(2)^2\over a(0)^2}- {a(0)^2\over a(1)^2}-{a(1)^2\over a(2)^2} \right)}\]
and similarly for general $n$. The  action on closer inspection can be viewed as a scalar field action $\sum \rho\Delta_{\Z_n}\rho$ for the postive-valued function $\rho=R_+(a)/a$ and $\Delta_{\Z_n}$ the standard finite-difference Laplacian. One can quantise the theory via $\rho$ or via the relative fluctuations $b=a/A$ where $A$ is the geometric mean of the $a(i)$, see\cite{ArgMa1}. Here, however, we just follow the straight theory. The results depend on $G$, which unlike the previous model is dimensionless and can't be scaled away. On the other hand, the dependence on $L$ is exact because we could move to scale-invariant variables $\bar a=a/L$ and then compute the integrals independently of $L$. Figure~\ref{Z3figs} shows some plots and one has, e.g. for $G=2$, 
\[ \<a(i)\>= 0.234 L,\quad \<a(i)a(j)\>=L^2 \begin{cases} 0.403 & i\ne j\\ 0.435 & i=j\end{cases},\quad  {\Delta a(i)\over \<a(i)\>}={\sqrt{
\<a(i)^2\>-\<a(i)\>^2}\over \<a(i)\>}=2.64,\]
which is similar to \cite{ArgMa1} and to which we now add the curvature
\[ \<R(i)\>=-{0.318 \over L}.\]
We see that the expectation values correspond to a regular polygon as the ground state,  but with nonzero standard deviation or `quantum fuzziness'  and nonzero curvature as quantum gravity effects (the regular polygon having zero curvature in QRG).  There is no need for a cutoff $\eps$ and these calculations do not noticeably change if we introduce a small $\eps$. 

The behaviour for $G\to 0$ can be analysed analytically as follows. The exponent the integrand of $Z$ is strictly negative other than zero at $a(0)=a(1)=a(2)$. Hence, we introduce new variables $a(0)=a, a(1)=a(1+x), a(2)=a(1+y)$ and when $G$ is small, only $x,y$ near to zero will contribute. Hence we can replace the exponent effectively by a quadratic function of $(x,y)$ and can then do the $x,y$ integrations as Erf functions. Up to a constant we obtain
\[ Z\sim  G\int_0^L a^2\extd a= {GL^3\over 3}\]
as $G\to 0$. Similarly with insertions of powers of $a(0)$, and then by symmetry, we have
\[ \<a(i)^m\>\to {3 L^m\over m+3},\quad \<a(i)\>\to {3 L\over 4},\quad {\Delta a(i)\over \<a(i)\>}\to {1\over\sqrt{15}}\]
which is consistent with the plots at $G=0.01$ (numerical integration at this value of $G$ is already unreliable and we cannot say much more than this). This is not classical behaviour but if we make the analogous change of variables for the $n$-gon then one similarly has
\[ \<a(i)^m\>\to {n L^m\over n+m},\quad {\Delta a(i)\over \<a(i)\>}\to {1\over\sqrt{n(n+2)}}\]
so that we will see classical behaviour in the limit of a circle at large $n$. Meanwhile, for the curvature of the triangle, we have  
\[ \<R(i)\>\sim -.699 {G\over L}\]
verified numerically for $G=0.01$ and consistent with $G\to 0$ asymptotic estimates for all $n$ along the same lines as above. This again fits the weak gravity limit where the regular $n$-gon has zero curvature. 

At the other extreme, i.e. for the strong gravity limit, we have
\[  Z\to L^3,\quad \<a(i)\>\to {L\over 2},\quad\<a(i)a(j)\>\to L^2\begin{cases}1/3 & i=j\\ 1/4 & i\ne j\end{cases},\quad{ \Delta a(i)\over \<a(i)\>}\to {1\over \sqrt{3}}\]
as $G\to\infty$. These limits are the same as working in the theory where $G=\infty$ and we just drop the exponential in the integrand, i.e. work with the rather trivial
\[ Z_0=\int_\eps^L\prod_i \extd a(i)= (L-\eps)^n=L^n(1-{1\over z})^n,\quad  z:={L\over\eps},\]
which is, moreover, the same for any graph with $n$ edges (it does not need to be an $n$-gon). Clearly, for powers of the metric other than $m_i=-1$,
\[ \< a(0)^{m_0}\cdots a(n-1)^{m_{n-1}}\>=\prod_i \left(L^{m_i}{[m_i+1]_{1\over z}\over m_i+1}\right);\quad [m]_{1\over z}:={1-{1\over z^m}\over 1-{1\over z}}\]
in which, for positive powers as above, we can just take $\eps=0$ or $z=\infty$. Here, expectation values are given by inserting the observable into the integral for $Z_0$  (i.e., integrating it) and dividing by $Z_0$.

The curvature, however, does depend on the graph and in the $Z_0$ theory for the $n$-gon, we do need the regulator $\eps$ to be able to do the integrals. The result, independently of $n$, is 
\[ \<R(i)\>=-{1\over L}\left({z+1\over 2 }+{\ln({1\over z})\over 1-{1\over z}}\right)\sim -{1\over 2\eps}\]
 if we expand for large $z$. This now depends on how we take the joint limit $\eps\to 0$ as $G\to\infty$. For example, setting $\eps=L/(4\sqrt{G})$ would give $\<R(i)\> L /\sqrt{G}\sim - 2$, while another choice  gives a better fit to 
 \[ \<R(i)\> L\sim -0.8 \sqrt{G\ln(G)}\]
 as shown in the figure (this is a good fit till about  $G=10^{4}$, the actual formula is likely to have several terms as we saw in the fuzzy sphere deep quantum gravity phase). This gives a flavour of how the divergence of $\<R(i)\>$ as $G\to\infty$ can be modelled in the limit $Z_0$ theory.

We won't give any details for the Lorentzian square because a new treatment extending \cite{Ma:squ}  will be given in  \cite{BliMa}. Suffice it to say that we see this as the group $\Z_2\times \Z_2$ (and numbered the vertices accordingly), which leads to a different $\Omega^2$ from the case of the 4-gon, as the natural left-invariant basis of 1-forms are different. This time there is a moduli of QLCs with an angle parameter $\theta$ which does not, however, enter the Einstein-Hilbert action
\[ S_g=(a_{00}-a_{01})^2({1\over a_{00}}+{1\over a_{01}})-(b_{00}-b_{10})^2({1\over b_{00}}+{1\over b_{10}})\]
obtained from the Ricci scalar curvature  $R$ and measure $\mu=ab$ built from the metric.  We again cut off metric values at scale $L$ and have non-zero $\<a\>$ and $\Delta a/\<a\>$ for the metric and a non-zero limit for these as $G\to \infty$, much as in our models above. Moreover, for the Ricci scalar at the different vertices,
\[ \<R(00)\>=\<R(11)\>={9\imath G\over 8 L^2}(1+\cos(\theta)),\quad  \<R(01)\>=\<R(10)\>={9\imath G\over 8 L^2}(1-\cos(\theta)).\]
which is imaginary.  The coupling constant $G$ as in (\ref{QG}) for this model has dimensions of length${}^2$, the same as $L$. Hence  $\bar G= G/L$ is comparable to the dimensionless $G$ for the polygon, after which we see that the result is analogous to the weak gravity limit there. 

\section{How elementary particles could emerge from quantum gravity}\label{secKK}

It is an old idea of Kaluza and Klein that gravity on a product spacetime $M\times K$ where $K$ is a compact Riemannian manifold with isometry group $G$ includes as certain modes within it Yang-Mills gauge theory on $M$ with group $G$ and gravity on $M$, see \cite{KK} for a review. Meanwhile, a scalar field on $M\times G$ appears as an infinite tower of fields with different masses coupled to the gauge field. This raises the hope of thinking of gauge fields as just part of a higher-dimensional gravity theory, but does not work too well in practice. For one thing, the metric on the product has to be very special, namely with coefficients that are constant on $K$ in the Killing form basis (the `cylinder ansatz'), i.e. this is a rather special ansatz and a way of looking at things but lacks explanatory power. Meanwhile, Connes and Chamsedine \cite{ChaCon} had the idea to replace $K$ by a finite noncommutative geometry $A_f$. Here $C^\infty(M)$ is replaced by $C^\infty(M,A_f)$ with values in $A_f$ and they were able to present the Standard Model of particle physics as resulting from a particular choice of $A_f$ and a particular `spectral triple' or abstract Dirac operator associated to it. This again was a particular product ansatz but did produce a postdiction for the Higgs mass and a novel mass relation among some matter fields. All of this was known for almost 20 years but only now do we have the tools to revisit the ideas using QRG.

What one finds \cite{LiuMa3} when $A_f$ is the fuzzy sphere is that everything now conspires to perfectly express the {\em full content} of gravity on $C^\infty(M)\tens A_f$ as exactly:
\begin{enumerate}
\item gravity on $M$ with its Einstein-Hilbert action for a field $\tilde g$.
\item $SU_2$ Yang-Mills gauge theory on $M$ for a field $\tilde A$ but slightly generalised in that $||\tilde F||^2$ uses a field $h_{ij}$ to contract the $su_2$ indices of the curvature $\tilde F$ of $\tilde A$.
\item $h_{ij}$ as a positive-matrix-valued Liouville field on $M$ which at each point is the metric $\ch=h_{ij}s^i\tens s^j$ on the fuzzy sphere at that point. Its Liouville potential is the Ricci scalar of $\ch$ on the fuzzy sphere.
\end{enumerate}
Specifically, the quantum metric on $C^\infty(M)\tens A_f$ is {\em forced} in local coordinates $x^\mu$ on $M$ to have the general form\cite{ArgMa4}
\[ \cg=g_{\mu\nu}\extd x^\mu \tens\extd x^\nu+ A_{\mu i}(\extd x^\mu\tens s^i+ s^i\tens A_\mu)+ h_{ij}s^i\tens s^j\]
and \cite{LiuMa3} proves that there is then a unique QLC, computes its Ricci scalar and then identifies its integral as the action for certain fields $\tilde g_{\mu\nu},\tilde A_{\mu i}$ built from $g_{\mu\nu},A_{\mu i},h_{ij}$ above.  Here, integration on the fuzzy sphere is taken to be the map that sends an element to its spin 0 component in the expansion of $A_f$ into $SU_2$ representations according to its action as orbital angular moment on the fuzzy sphere. 

Next, we saw in Section~\ref{secbaby} that in quantum gravity on the fuzzy sphere, the expected value of the metric is the round metric $h_{ij}=h\delta_{ij}$ (since all the $\lambda_i$ have the same value independently of $i$) and, moreover, $h$ might be expected to be a constant (if we regularise and renormalise uniformly over spacetime). Hence the first approximation to what we see on $M$ is exactly $SU_2$-Yang-Mills with $h$ now matched to whatever coupling constant is desired for that. In \cite{LiuMa2} as well as in the original works of Kaluza and Klein where the fibre is a circle, one needs the fibre metric scale factor to be $h\approx (23\lambda_P)^2$ to match the coupling constant for electromagnetism. If we wanted a similar $\<h\>$ in Section~\ref{secbaby} then we would then need $L=530$ in Planck area units, so well in the $L>>6$ phase. For the electroweak force, the effective strength from particle scattering depends on the energy scale at which you measure it, but would appear to be again in the large $L$ phase. If such ideas applied to the strong force (which is not immediately the case due to a different gauge group), this again depends on the energy scale and is now stronger for lower energies due to asymptotic freedom. This at lower energies  could be 100 times stronger than electromagnetism (i.e., a strong fine structure constant of $\alpha_S\approx 0.7$, say). This translates to $\<h\>$ and hence $L\approx 5$ which would be in the other phase of the theory if results for the relevant model were comparable to ours for the fuzzy sphere. This would need to be explored for more relevant quantum geometries (with $SU_3$ symmetry) as fibre. 

It also remains, even for the fuzzy sphere, to see how fermions (needed for matter fields) on the product appear as certain multiplets on $M$. For scalar fields, the fuzzy sphere as studied is not finite-dimensional but for $\lambda_P=1/n$ for an integer $n$, it has a natural quotient as a matrix algebra. In this case, scalar fields decompose as a finite multiplet of fields on $M$ of different masses, again fitting better to the Standard Model and providing a route to understanding the `generations problem' (i.e. why the electron, its neutrino and the up/down quarks appear to be repeated two more times with particles of different mass but otherwise identical). In this context, the $SU_2$ gauge field would not be the electroweak one but an approximate flavour symmetry that mixes the generations, for example a triplet from the spin 1 representation as discussed in \cite{LiuMa3}. 

\section{Quantum geodesics}\label{secgeo}

If quantum spacetimes have no points, then it is hard to imagine that they have geodesics. These can nevertheless be addressed by taking a fluid-mechanics like point of view. Thus, consider not one particle but a dust of particles with density $\rho$ on a (pseudo)Riemannian manifold $M$, where each particle moves on a geodesic. At first sight one might think that the tangent vectors of all these particles fit together to form a vector field $X$ that also evolves in time. This will need to obey the {\em geodesic velocity equation} 
\[ \dot X+ \nabla_X X=0\]
where the dot means with respect to the geodesic `time' parameter, which we will denote by $s$ to avoid confusion in the case where $M$ is spacetime. Moreover,
\[ \dot\rho= - X(\extd \rho)+ {\rm div}(X)\rho\]
is the continuity equation in the fluids case, but we will refer to it as the {\em density flow equation}. What you may not realise, however is that $X$ {\em cannot be determined by $\rho$}, but meanwhile the geodesic velocity equation can be solved without reference to any particles at all! Hence, if we have access only to densities and not particles themselves, we have to rip apart the usual concept of a geodesic and re-assemble it in reverse order: we think of the geodesic vector field $X$ as a physical quantity in its own right, and then after solving for $X$ we can evolve any initial density to later time via the density flow equation. This new field $X$ captures a lot of geometry, for example\cite{BegMa:cur}
\[ {D\over D s} {\rm div}(X)=- ||\nabla X||^2 - {\rm Ricci}(X,X)\]
where $D\over Ds$ is the convective or co-moving derivative. This directly shows the role of the Ricci tensor in controlling how a fluid expands or contracts in free-fall. 

Next, we go one stage further and take a leaf out of quantum mechanics, replacing $\rho$ by a complex amplitude $\psi$ with $\rho=|\psi|^2$. We replace the density flow by the {\em amplitude flow}
\[ \dot\psi+X(\extd \psi)+\psi {1\over 2} {\rm div}(X)=0.\]
This is equivalent for $\psi$ real and positive but for other, e.g., complex, values it opens up the possibility of interference phenomena which remain to be explored, for example around a regular classical black hole. The `quantum mechanical interpretation' here applies to the observer with time $s$, which is external to the system. In GR, geodesics entail a so-called `God's eye view' and that eye is now being upgraded to a quantum mechanical language. 

\begin{figure}
\[ \includegraphics[scale=.97]{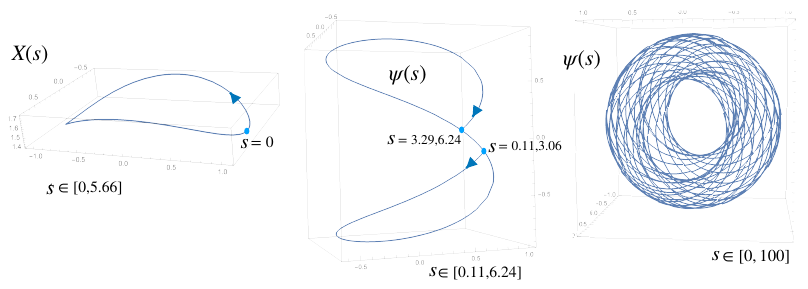}\]
\caption{\label{figgeo} Quantum geodesic on the fuzzy sphere. We restricted to wave functions of the form $\psi=\psi_i x^i$ with $\psi_i$ real. Unitary evolution ensures $\vec\psi$ stays on a sphere, but there are two disks into which the geodesic never enters.  Image adapted from \cite{BegMa:cur}.}
\end{figure}

At this point, everything extends to a general QRG. Given $(A,\Omega^1,\extd)$, we let $X\in \chi:={}_A\hom(\Omega^1,A)$ be a left vector field (a left-module map for the action of $A$ from the left). A QLC on $\Omega^1$ induces a right handed connection $\nabla_\chi:\chi\to \chi\tens\Omega^1$ and the geodesic velocity equation in the simplest form becomes\cite{BegMa:geo}
\[ \dot X+ [X,{1\over 2}{\rm div}(X)]+ (\id\tens X)\nabla_\chi X=0,\]
where ${\rm div}(X)$ is the divergence of $X$ defined ideally via the QLC. In the simplest case, we let $\psi\in A$ (or rather in a completion of this where $\int \psi^*\psi<\infty$ with respect to a chosen integration) and take the same equation as above extended to this, being careful about the order. This works in the simplest case where $\int$ is a twisted trace and $\int {\rm div}(X)=0$ for all $X$. We refer to \cite{BegMa:cur} for details. Under these assumptions, one can show that
\[ {\extd \over\extd s}\int \psi^*\psi=0\]
so that we can in principle normalise $\rho=\psi^*\psi$ to $\int\rho=1$. Here $\rho$ is an evolving positive operator not a real density, but one still has a probabilistic picture with expectation value 
\[ \<a\>={\int \psi^* a\psi\over \int \psi^*\psi}\]
in state $\psi$. An example of a quantum geodesic on the fuzzy sphere with metric $g_{ij}={\rm diag}(4,3,1)$ is given in Figure~\ref{figgeo} taken from \cite{BegMa:cur}, where $X$ is solved in terms of elliptic Jacobi functions,
\[X^1(s)=-\tfrac{1}{\sqrt{2}}\, \text{sn}(s|-\tfrac{1}{2}),\quad X^2=\text{cn}(s|-\tfrac{1}{2}),\quad X^3=\sqrt{2+\text{sn}^2(s|-\tfrac{1}{2})},\]
where $X=f_iX^i$ and $f_i$ are  a dual basis to the $s^i$ in Section~\ref{secbaby}. This is an easy case where each $X_i$ is constant on the fuzzy sphere, and likewise $\psi$  is solved in a easy case where it is linear in the coordinates. 

This machinery has the power to give coordinate-free predictions in quantum spacetime models. So far, this has only been explored in first approximation (without consideration of the functional analysis) but two take-aways on the spacetime (\ref{bicmodel}) are\cite{LiuMa1}

\begin{enumerate}
\item  An initial $\psi$ which is a Gaussian bump at a point in spacetiime and then evolves (approximating the worldline of a point particle) gets quantum corrections which are inverse to the Gaussian width, i.e. a perfect point source is likely forbidden by quantum gravity due to infinite corrections mediated  by the quantum spacetime. 

\item An initially real-valued $\psi$ typically gets complex quantum corrections, i.e. does not stay real. Hence quantum gravity will lead to the kind of interference effects mentioned above.
\end{enumerate}

This new technology requires further study and in particular could be applied to noncommutative black hole models and expanding universe (FLRW) cosmological models. Models of these within QRG appeared in \cite{ArgMa2} and one interesting feature is a {\em dimension jump} where, if we replace the sphere at each $r,t$ by a fuzzy sphere and look for spherically symmetric static solutions with $r,t$ classical, we end up with a metric similar to the 5D Tangherlini black hole. We can also replace the sphere at each point by $\Z_n$ or its limit of a circle with its limiting 2-dimensional calculus. Ditto for cosmological models, an expanding $S^3$ at each time can be replaced by an expanding fuzzy sphere with the same Friedmann expansion equation as the Standard Cosmological Model.  Dispersion relations have been long-conjectured to be modified in quantum spacetime but these could now be determined by looking at quantum geodesic flows. 

Finally, quantum geodesics should also provide a new framework to see particle creation or `Bekenstein-Hawking radiation'. This has already been demonstrated on the integer line graph (a 1-dimensiosional lattice) in \cite{Ma:haw} as follows. We assume a constant flat metric $a(i)$ at far left (large negative $i$) and again at far right (large positive $i$) and a varying metric in a region around $i=0$. Now, we can solve the wave equation $(-\Delta+m^2)\phi=0$ starting with boundary conditions matching a plane wave at far left for the metric there and then solving the recursion entailed in the wave equation to extend $\phi$ to all $i$. At large $i$ this will typically no longer be a plane wave there but one can match it to ones with different momenta components. Usually, particle creation is couched in the language of quantum mechanics and the vacuum at far left being transported to the far right as no longer a vacuum state, but the above should also show the effect. A question for the future is whether we can see similar effects with quantum geodesics. Here, one approach is to follow \cite{BegMa:geo} where the `Klein Gordon flow' $\dot\psi={\imath\hbar\over 2m}\Delta\psi$ can be viewed as an amplitude flow for a certain geodesic velocity field defined with respect to a generalised (not symmetric) quantum metric on the relevant Heisenberg algebra. This was on ordinary spacetime with a background Maxwell field but the ideas could be adapted to any QRG. 

\section{Concluding remarks}\label{secrem}

The quantum spacetime hypothesis, which we explored, is agnostic as to the `true' theory of quantum gravity, in that several different approaches ranging from string theory to loop quantum gravity and spin networks suggest that this is closer to the real world than a continuum due to quantum gravity effects. In this case,  however, it would be more consistent to construct quantum gravity on a quantum spacetime, and consistently with it, than to build quantum gravity on a continuum. In this respect, approaches built on a continuum such as supergravity and string theory, while very interesting, do not directly address the problem at hand.  

By contrast, letting the coordinate algebra be noncommutative and/or finite dimensional, as seen in our baby quantum gravity models, turns the infinities that plague  straight continuum quantum gravity into ordinary divergences that are more reasonable to handle, even if we still need to renormalise. We now return to our check-list Q1-Q5 and ask what we have learned from such models and the related formalism of QRG so far.  For Q1, we have seen some small hints in Section~\ref{secgeo} that point-sources are not physical due to quantum corrections. This needs to be looked at further and meanwhile is sidestepped for graphs, where the entire space is discretised so that the issue does not arise. For Q5, we saw in Section~\ref{secKK}  a mechanism for how multiplets of particles transforming under a local gauged symmetry could naturally arise from gravity on a product {\em provided}  we take the fibre to be a QRG. Taking $A$ to be highly noncommutative in the sense of trivial centre was key here for the mechanism.  For Q3, baby quantum gravity models in principle allow one to `count' everything explicitly and there is no particular problem doing the computations. Moreover, optimal transport theory on graphs takes a similarly  probabilistic approach but has well-developed links with entropy and quantum information\cite{Lott}. We also know from QRG models that  thermalisation of the vacuum can be seen as a general feature of transition through a region of varying metric\cite{Ma:haw} and how one can do calculations. Hence, this question appears on track but requires more development. Q4 relates to the nature of quantum measurement and here, rather than solving anything, I would like to say we may need a paradigm shift. Quantum geodesics, which apply quantum methods to the process of observing geodesic flow,  could be a step in the right direction. 

Finally, for Q2, we do have indications that dark energy or the cosmological constant could arise from quantum spacetime. First, we saw in the baby quantum gravity models a nonzero relative uncertainty in the metric, which suggests some kind of vacuum energy. For a discrete circle modelled by $\Z_n$, we saw that this goes as $1\over n$ or the relative variance $1\over n^2$. One might speculate that the energy density corresponding to this should be the Planck density (due this being a quantum gravity effect) times the relative variance. We have no strong argument for this, but we note that  if we crudely model the universe as a discrete circle $\Z_n$ then we might want $n=5.5 \times 10^{61}$ as the size of the Universe in Planck units, and in this case
\[ {{\rm Planck\ density}\over n^2}\approx {5.1\times 10^{93}\over  (5.5 \times 10^{61})^2}\approx  2\times 10^{-29}{\rm  g/cm}^3 \]
compares very well with the observed density of the Universe of $9\times 10^{-30}$, most of which is believed to be dark energy. Or from another angle, for a finite chain $\bullet$--$\bullet$-$\cdots$ - $\bullet$ with $n$ nodes, one finds\cite{ArgMa3} that the natural QRG is q-deformed with $q=e^{\imath \pi\over n+1}$  compared to an infinite lattice. But in Euclideanised 2+1 quantum gravity, q-deformation corresponds to introduction of a cosmological constant $\Lambda=\lambda_c^{-2}$ via $q=e^{\lambda_c\over\lambda_P}$. Equating these (without the $\imath$) with our above value of $n$ gives $\lambda_c\sim \lambda_P {n\over\pi}\sim 3 \times  10^{26}$m or cosmological constant $10^{-53}$m${}^{-2}$, compared to the observed value of $10^{-52}$m${}^{-2}$. Thus, while these arguments are both crude and speculative, they could be seen as supporting the idea that the cosmological constant arises from the noncommutative nature of spacetime, which was discrete in the above discussion. Note that from the point of view of 2+1 quantum gravity, the introduction of the cosmological constant suggests a q-deformation of (\ref{spinmodel}) to the quantum group $U_q(su_2)$ as the relevant model quantum spacetime\cite{MaSch}. This means that the curved $SU_2$ momentum space dual to (\ref{spinmodel}) becomes the quantum group $C_q(SU_2)$, i.e. the cosmological constant from this point of view leads to noncommutative momentum space, in addition to changing the spacetime further. 
 
To proceed from such hints to actual calculations, a better measure of the energy density in the gravitational field is $R\over 16 \pi G_N$ where $G_N$ is the actual Newton constant (in our baby models $G$ had a similar role but was of whatever dimensions was needed for a dimensionless exponent, for example of dimensions length${}^2$ for the fuzzy sphere).  We are interested in quantum-gravity induced curvature corrections and for the fuzzy sphere our next-to-leading order result can be written as $2\<R\>/L^2$. If we match the curvature there with our observed curvature of order ${1/\lambda_c^2}= \Lambda$  then the quantum-gravity induced energy density is
\[ {\Lambda c^2 \over 8 \pi G_N L^2}\approx { 10^{-52} c^2\over 1.7\times 10^{-12} L^2} \approx {5.3\times 10^{-30} \over L^2}{\rm  g/cm}^3\]
where $c$ is the speed of light and the final expression is converted from cubic metres to cubic cm. Since we assumed that $L>>6$, this is even smaller than we wanted but provides a proof of concept as to how this might work. The problem, however, is that the dimensionless parameter $L$ enters here as a regulator and its physical interpretation is not fully understood. We could, for example, replace it to leading order by $L=\<\lambda_i\>=\<\lambda_i^{phys}\>/G=\lambda^2_U/G$ for $\lambda_U$  the size of the Universe, but the question would still be the value of $G$ at this scale. The obvious choice $G=\lambda_P^2$ leads to an extremely small answer and  we would need something rather closer to the other end  $G=\lambda_U^2$. Alternatively, we could be in the $L<<6$ phase of the model and in this case have to contend with a divergence of $\<R\>$ in terms of a regulator $\eps$ and its more significant renormalisation. There is only so much one can learn about the real world by analogy here, but this gives a flavour of some issues for further work towards more appropriate  models. 

Turning now to the mathematical side, the measure $\mu$ in constructing the baby quantum gravity models was chosen somewhat to taste and a more systematic approach to this is needed. Some results in this direction are suggested by the compatibility of the integral with divergence as in the theory of quantum geodesics\cite{BegMa:cur}. Different choices will change the dimension of $G$ and the flavour of the model. We will also need to understand better and perhaps modify the construction of the Ricci tensor. Both issues will need to be informed by a theory of variational calculus in noncommutative geometry, which is currently lacking and which is needed to connect the path integral approach adopted so far, to an operator level `Hamiltonian' approach. Variational calculus in turn needs either a proper understanding of the Hopf algebroid of differential operators on a differential algebra or, on the dual side, of its jet bundle. Both of these are an active area of development at the time of writing, with some initial work\cite{Gho,MaSim1, Flood}. A further clue is that quantum geodesics themselves should be understood in terms of variational calculus as is the case classically.


\begin{thebibliography}{99}

\bibitem{AmeMa} G. Amelino-Camelia and S. Majid, Waves on noncommutative spacetime and gamma-ray bursts, Int. J. Mod. Phys. A 15 (2000) 4301--4323

\bibitem{ArgMa1}J. Argota-Quiroz and S. Majid, Quantum gravity on polygons and $\Bbb R \times \Bbb Z_n$ FLRW model, Class. Quantum Grav. 37 (2020) 245001

\bibitem{ArgMa2} J. Argota-Quiroz and S. Majid, Fuzzy and discrete black hole models, Class. Quant. Grav. 38 (2021) 145020

\bibitem{ArgMa3} J. Argota-Quiroz and S. Majid, Quantum Riemannian geometry of the discrete interval and q-deformation, J. Math. Phys. 64 (2023) 051701

\bibitem{ArgMa4} J. Argota-Quiroz and S. Majid, Quantum gravity on finite spacetimes and dynamical mass, Corfu Summer Institute 2021: School and Workshops on Elementary Particle Physics and Gravity, PoS (2022) 210

\bibitem{FreMa}L. Freidel and S. Majid, Noncommutative harmonic analysis, sampling theory and the Duflo map in 2+1 quantum gravity, Class. Quant. Gravity 25 (2008) 045006

\bibitem{Beg:geo}E.J. Beggs, Noncommutative geodesics and the KSGNS construction, J. Geom. Phys. 158 (2020) 103851

\bibitem{BegMa:rie} E.J. Beggs and S. Majid, *-Compatible connections in noncommutative Riemannian geometry, J. Geom. Phys. 61 (2011) 95-124

 \bibitem{BegMa:gra} E.J. Beggs and S. Majid, Gravity induced by quantum spacetime, Class. Quant. Grav. 31 (2014) 035020 (39pp)


\bibitem{BegMa}E.J. Beggs and S. Majid, Quantum Riemannian Geometry, Grundlehren der mathematischen Wissenschaften, vol. 355, Springer (2020) 809pp.

\bibitem{BegMa:geo}E.J. Beggs and S. Majid, Quantum geodesics in quantum mechanics, J. Math. Phys. 65 (2024) 012101

 \bibitem{BegMa:cur}E.J. Beggs and S. Majid, Quantum geodesics and curvature, Lett. Math. Phys. (2023) 113:73
 
 \bibitem{BliMa} S. Blitz and S. Majid, Quantum gravity on a square and FLRW corrections, {\em in preparation}

\bibitem{ChaCon}A. Chamsedine and A. Connes, Why the Standard Model, J. Geom. Phys. 58 (2008) 38-47

\bibitem{Con}A. Connes,   Noncommutative Geometry,  Academic Press, Inc., San Diego, CA, 1994

\bibitem{DFR}S. Doplicher, K. Fredenhagen and J. E. Roberts, The quantum structure of spacetime at the Planck scale and quantum fields, Commun. Math. Phys. 172 (1995) 187--220

\bibitem{Dri}V.G. Drinfeld, Quantum Groups. In {\em Proc. ICM 1986}, AMS, Rhode Island. 


\bibitem{DVM}
M. Dubois-Violette and  P.W.\ Michor, Connections on central bimodules in 
noncommutative differential geometry, J.\ Geom.\ Phys.\ 20 (1996) 218--232

\bibitem{Flood} K.J. Flood, M. Mantegazza, H. Winther, Jet functors in noncommutative geometry, arXiv:2204.12401 (math.qa)

\bibitem{Gho} A. Ghobadi,  Hopf algebroids, bimodule connections and noncommutative geometry, arXiv:2001.08673 (math.qa)
 


\bibitem{Hoo}G. 't Hooft, Quantization of point particles in 2+1 dimensional gravity and space- time discreteness, Class. Quant. Grav. 13 (1996) 1023



\bibitem{LirMa1} E. Lira-Torres and S. Majid, Quantum gravity and Riemannian geometry on the fuzzy sphere, Lett. Math. Phys. (2021) 111:29

\bibitem{LiuMa1}C. Liu and S. Majid, Quantum geodesics on quantum Minkowski spacetime,  J. Phys. A 55 (2022) 424003

\bibitem{LiuMa2} C. Liu and S. Majid, Quantum Kaluza-Klein theory with $M_2(\C)$, JHEP (2023) 102

\bibitem{LiuMa3} C. Liu and S. Majid,  Yang-Mills fields from fuzzy sphere quantum Kaluza-Klein model arXiv:2312.14128 (hep-th)

\bibitem{Lott} J. Lott and C. Villani, Ricci curvature for metric-measure spaces via optimal transport. Annals Math. (2009) 903--991

\bibitem{Luk}J. Lukierski, H. Ruegg, A. Nowicki and V.N. Tolstoy, q-deformation of Poincar\'e algebra, Phys. Lett. B 264 (1991) 331

\bibitem{Mad}J. Madore, The fuzzy sphere, Class. Quantum Grav. 9 (1992) 69--88

\bibitem{Ma:pla}S. Majid, Hopf algebras for physics at the Planck scale, Class. Quant. Grav. 5 (1988) 1587--1607

\bibitem{Ma:squ} S. Majid, Quantum gravity on a square graph, Class. Quantum Grav 36 (2019) 245009

\bibitem{Ma:haw}
S. Majid, Quantum Riemannian geometry and particle creation on the integer line, Class. Quantum Grav 36 (2019) 135011 

\bibitem{MaRue}S. Majid and H. Ruegg, Bicrossproduct structure of the $\kappa$-Poincare group and non-commutative geometry, Phys. Lett. B. 334 (1994) 348--354

\bibitem{MaSch}S. Majid and B. Schroers, q-Deformation and semidualisation in 3D quantum gravity, J. Phys A 42 (2009) 425402

\bibitem{MaSim1}S. Majid and F. Simao, Quantum jet bundles, Lett. Math. Phys. (2023) 113:120


\bibitem{Mou} J.\ Mourad, Linear connections in noncommutative geometry, Class.\ Quant. 
Grav.\ 12 (1995)  965--974

\bibitem{KK}J.M. Overduin and P.S.  Wesson, Kaluza-Klein Gravity. Physics Reports. 283 (1997)  303--378

\bibitem{SeiWit} N. Seiberg and E. Witten. String theory and non-commutative geometry. JHEP 09 (1999) 032





\end{thebibliography}
\end{document}